\documentclass[aps,prl,twocolumn]{revtex4-1}
\usepackage{amsmath}
\usepackage{amssymb}
\usepackage{graphicx}
\usepackage{epstopdf}
\usepackage[colorlinks=true]{hyperref}

\usepackage{physics}
\usepackage{mathrsfs}
\usepackage{comment}

\newcommand{\sectionn}[1]{{\textit{#1}}}
\renewcommand\({\begin{equation}}	% quick macro for equation with numbers
\renewcommand\){\end{equation}}

\begin{document}

\title{Energy storage in magnetic textures driven by vorticity flow}

\author{Dalton Jones}
\affiliation{Department of Physics and Astronomy, University of California, Los Angeles, California 90095, USA}

\author{Ji Zou}
\affiliation{Department of Physics and Astronomy, University of California, Los Angeles, California 90095, USA}

\author{Shu Zhang}
\affiliation{Department of Physics and Astronomy, University of California, Los Angeles, California 90095, USA}

\author{Yaroslav Tserkovnyak}
\affiliation{Department of Physics and Astronomy, University of California, Los Angeles, California 90095, USA}

\begin{abstract}
An experimentally feasible energy-storage concept is formulated based on vorticity (hydro)dynamics within an easy-plane insulating magnet. The free energy, associated with the magnetic winding texture,  is built up in a circular easy-plane magnetic structure by injecting a vorticity flow in the radial direction. The latter is accomplished by electrically induced spin-transfer torque, which pumps energy into the magnetic system in proportion to the vortex flux. The resultant magnetic metastable state with a finite winding number can be maintained for a long time because the process of its relaxation via phase slips is exponentially suppressed when the temperature is well below the Curie temperature. We propose to characterize the vorticity-current interaction underlying the energy-loading mechanism  through its contribution to the effective electric inductance in the rf response. Our proposal may open an avenue for naturally powering spintronic circuits and nontraditional magnet-based neuromorphic networks. 
\end{abstract}

\date{\today}
\maketitle

%%%%%%%%%%%%%%%%%%%%%%%%%%%%%%%%%%%%%%%%%%%%%%%%%%%%%%%%%%%%%%%%%%%%%%%%%

\sectionn{Introduction.}|The centerpiece of the global energy challenge today is a viable method for energy storage, whose key is to convert captured energy into forms that are convenient or economic for long-term storage. 
Commonly used forms of energy storage are based on chemical energy (lithium-ion batteries), gravitational energy (hydroelectric dam), thermal energy (molten salt), etc. 
Recent progress in the field of spintronics enables us to manipulate magnetic textures in numerous ways~\cite{spincurrent22, Yaroslavreview, Ochoa:2019aa}, which inspires  the possibility of storing energy in the exchange energy associated with topological magnetic textures~\cite{PhysRevLett.121.127701}.

Here, we propose a feasible scheme for energy storage in the topological magnetic winding texture of a magnetic insulator---a magnetic battery. The physical mechanism for charging or discharging is through vortex hydrodynamics. The ``phase-slip'' phenomenon in a spin superfluid~\cite{PhysRevLett.116.127201, *PhysRevB.93.020402} is known to reduce the phase winding of a one-dimensional system by $2\pi$, by sending a vortex through it. Vice versa, driving a vortex flow in the opposite direction will naturally build up the winding number, and hence the magnetic exchange energy. 

Although our system is limited in terms of energy density compared with the prevalent lithium-ion battery technology, our approach does have  a few advantages. First, magnetic systems are highly nonvolatile and endurable. Magnetic textures protected by nontrivial topological numbers, such as domain walls, vortices, and skyrmions, have already been employed in memory and logic devices~\cite{Allwood1688, Fert:2013aa, Parkin190, RevModPhys.80.1517}. Energy can be stored over an extremely long time scale, with essentially no degradation in charging and discharging cycles. 
Second, magnetic batteries can be naturally incorporated into spintronic circuits~\cite{Allwood1688, Chumak:2015aa, Khitun_2010, PhysRevX.5.041049}, neuromorphic platforms~\cite{7563364, Yueeaau8170, Torrejon:2017aa, Sengupta_2018, nuero}, and quantum-information processing tasks based on insulating magnets~\cite{PhysRevB.95.144402, PhysRevB.101.014416, PhysRevB.100.174407}, rendering coherent and low-dissipation operations based purely on spin dynamics. 
Third, common magnetic materials are environmentally friendly and the development of magnetic batteries is another possible avenue leading to the goal of clean energy.

\begin{figure}
\hfill\includegraphics[scale=0.23]{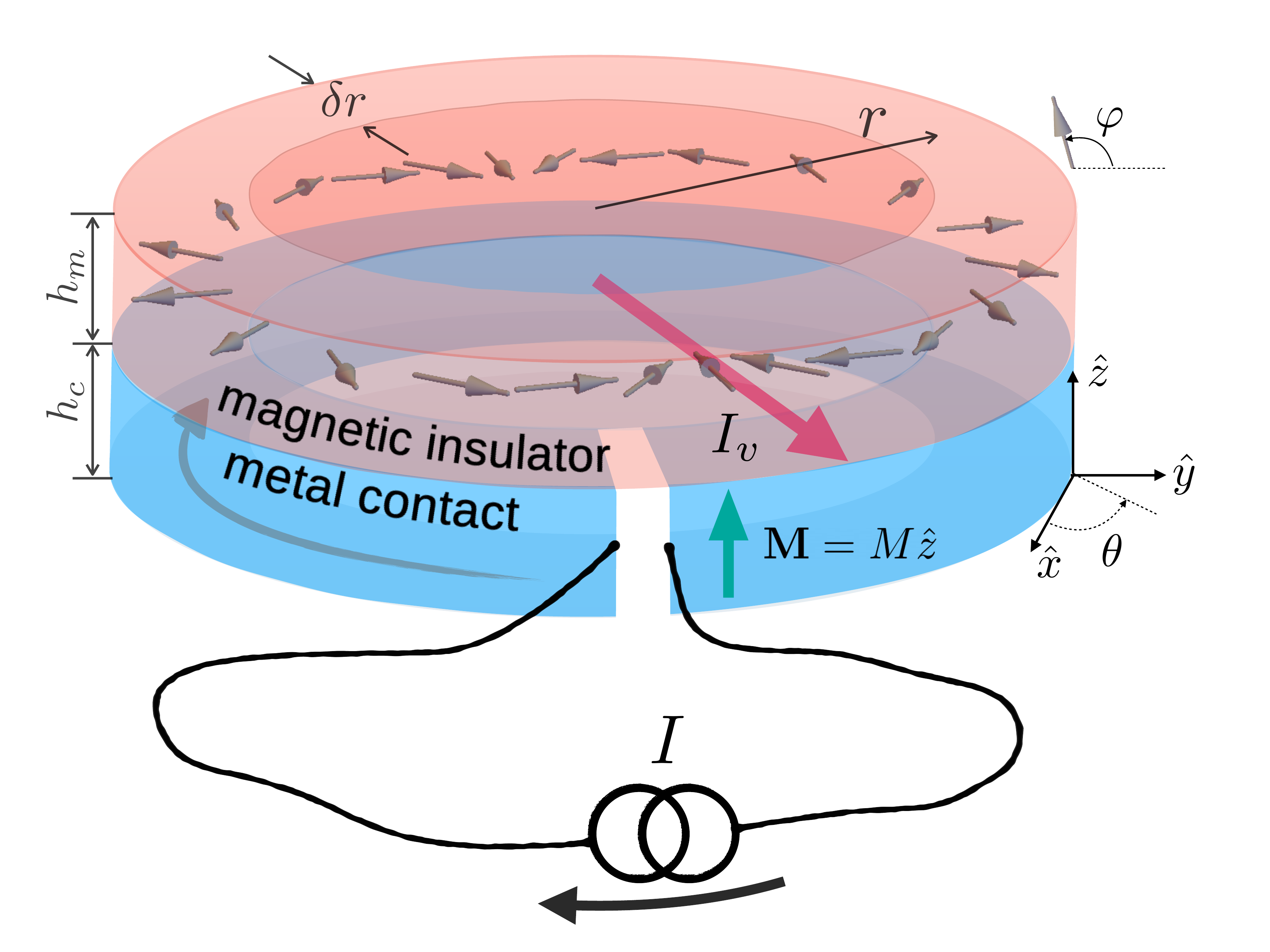}
\caption{The ring-shaped bilayer with a radius $r$, width $\delta r$, and heights $h_m$ for the magnetic insulator and $h_c$ for the metallic contact. The magnetic insulator has an easy-$xy$-plane anisotropy for the order parameter, whose in-plane orientation is parametrized by the azimuthal angle $\varphi$. $\theta$ is the polar angle for a real-space position along the ring. The (ferromagnetic) metal layer has a uniform magnetization $\vb{M}= M\hat{\mathbf{z}}$ and an azimuthal current $I$. The electric current induces a vortex flow $I_v$ in the radial direction, which builds up an azimuthal winding density $\partial_x \varphi$ of the magnetic order parameter. 
}
\label{fig1}
\end{figure}

\begin{figure}
\hfill\includegraphics[scale=0.23]{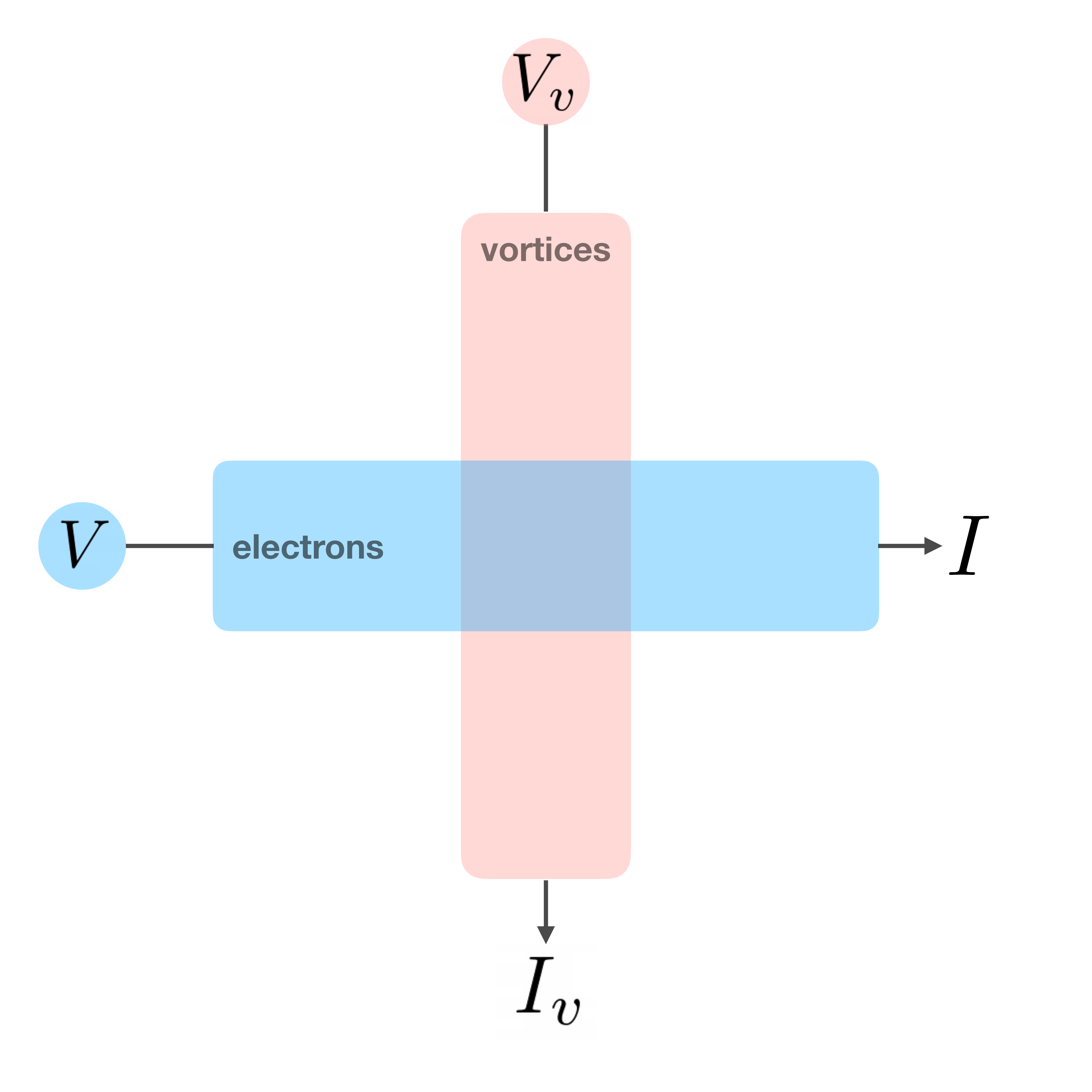}
\caption{Schematic in Fig.~\ref{fig1} shows two viscously coupled hydrodynamic entities: one is the electron flow $I$ and the other is the vortex flow $I_v$. This coupling is Magnus cross-like, meaning that a horizontal electron flow drives a vertical vortex flow and vice versa. 
}
\label{fig3}
\end{figure}

%%%%%%%%%%%%%%%%%%%%%%%%%%%%%%%%%%%%%%%%%%%%%%%%%%%%%%%%%%%%%%%%%%%%%%%%%
\sectionn{Central concept.}|To illustrate our main idea, we consider an annulus structure depicted in Fig.~\ref{fig1}. A thin-film easy-plane magnetic insulator is placed on top of a metal contact. The magnetic insulator can be ferromagnetic or antiferromagnetic, with an order parameter $\vb{n}(\vb{r},t)$, which is fluctuating in and out of the $xy$ plane. Its winding texture is described by the azimuthal angle $\varphi(\theta,t)$, where $\theta$ is the polar-coordinate angle. The metal annulus has a uniform magnetic order $\vb{M}= M\hat{\mathbf{z}}$.

We define the vorticity 3-current in $(2+1)$ dimensions within the thin-film magnetic insulator as
\( \mathcal{J}^\mu = \epsilon^{\mu\nu\rho} \hat{\mathbf{z}} \cdot ( \partial_\nu \vb{n} \times \partial_\rho \vb{n})/2\pi , \label{3} \)
which is carried by the magnetic texture~\cite{jivortex}. Here $ \epsilon^{\mu\nu\rho}$ is the Levi-Civita symbol (with the Einstein summation implied over the Greek indices $\mu=0,1,2\leftrightarrow t,x,y$). The current obeys a topological conservation law, $\partial_\mu \mathcal{J}^\mu=0$. The total vortex number in the bulk $\Omega$
\(\mathcal{N}=\int_\Omega dxdy\,\,  \mathcal{J}^0=\frac{1}{2\pi} \int_{\partial \Omega} d\vec{l} \,\, \vb{n}_{\|}^2 \, \vec{ \nabla} \varphi  \label{4}  ,\)
by Stokes theorem, is also the total winding number at the boundary $\partial \Omega$. Here $\vb{n}_{\|}$ is the easy-plane projection of the order parameter $\vb{n}$. We remark that this construction is true not only at the low-temperature regime, where $\mathcal{N}$ is integer-valued, but also applicable at high temperatures and the paramagnetic regime (even in the lattice limit \cite{quantumvortex}), where the vortex number is not quantized. 

To load the free energy associated with the magnetic winding texture, we operate the magnetic system near the Curie temperature (paramagnet regime) so that vortices and anti-vortices deconfine to form a two-dimensional, two-component plasma with finite vortex conductivity $\sigma_v$ \cite{jivortex}.  A constant electric current $I$ circulating in the magnetic metal contact (see Fig.~\ref{fig1})  energetically biases a radial vortex flow $I_v$~\cite{jivortex} based on symmetry analysis.  We articulate the detailed mechanism in later section. The electric current and vortex current are Magnus cross-coupled as shown in Fig.~\ref{fig3}.

Using this externally driven vortex flow, we are able to reverse the typical ``phase-slip'' process in superfluids~\cite{PhysRevLett.116.127201, PhysRevB.93.020402, Girvinbook,Halperinreview} and build up a finite order-parameter winding density $\partial_x \varphi$ in the magnetic insulator. The rate of change of the magnetic winding number and the intensity of the vorticity flow are related by the conservation law for the vortex 3-current \eqref{3}:
\(\text{d}\mathcal{N}/\text{d}t=I_v.  \)
As the winding number accumulates, the metastable magnetic configuration builds up a finite free-energy density and exerts a restoring force on the vortex flow, which decays exponentially and eventually vanishes when the restoring force by the winding texture balances the external drive. This type of process is analogous to the experimental proposal by Pearl~\cite{SCvort}, in which a magnetic screw rotating inside a superconducting cylinder is used to propagate vortices radially in order to increase the azimuthal superflow. In this system, the mechanical energy of the rotating magnetic screw is converted into the energy associated with the increased winding of the order parameter. Similarly, our system converts electrical energy into the exchange energy of the magnetic texture.

Tuning the temperature for our magnetic system well below the Curie temperature $T_c$ keeps the winding texture within plane, due to the easy-plane anisotropy, thus endowing it with a topological protection. In this regime, the conductivity of vortices  is  frozen and the unwinding process is exponentially suppressed. As a result,  the energy associated with the magnetic texture can be stored indefinitely in the absence of an external drive. To release the energy stored in the magnetic winding texture, we can simply raise the temperature to near $T_c$  and make use of the natural vortex flow in the ``phase-slip'' regime. The electromotive force from the vortex flow becomes the output voltage of the magnetic battery.

\begin{figure}
\hfill\includegraphics[scale=0.23]{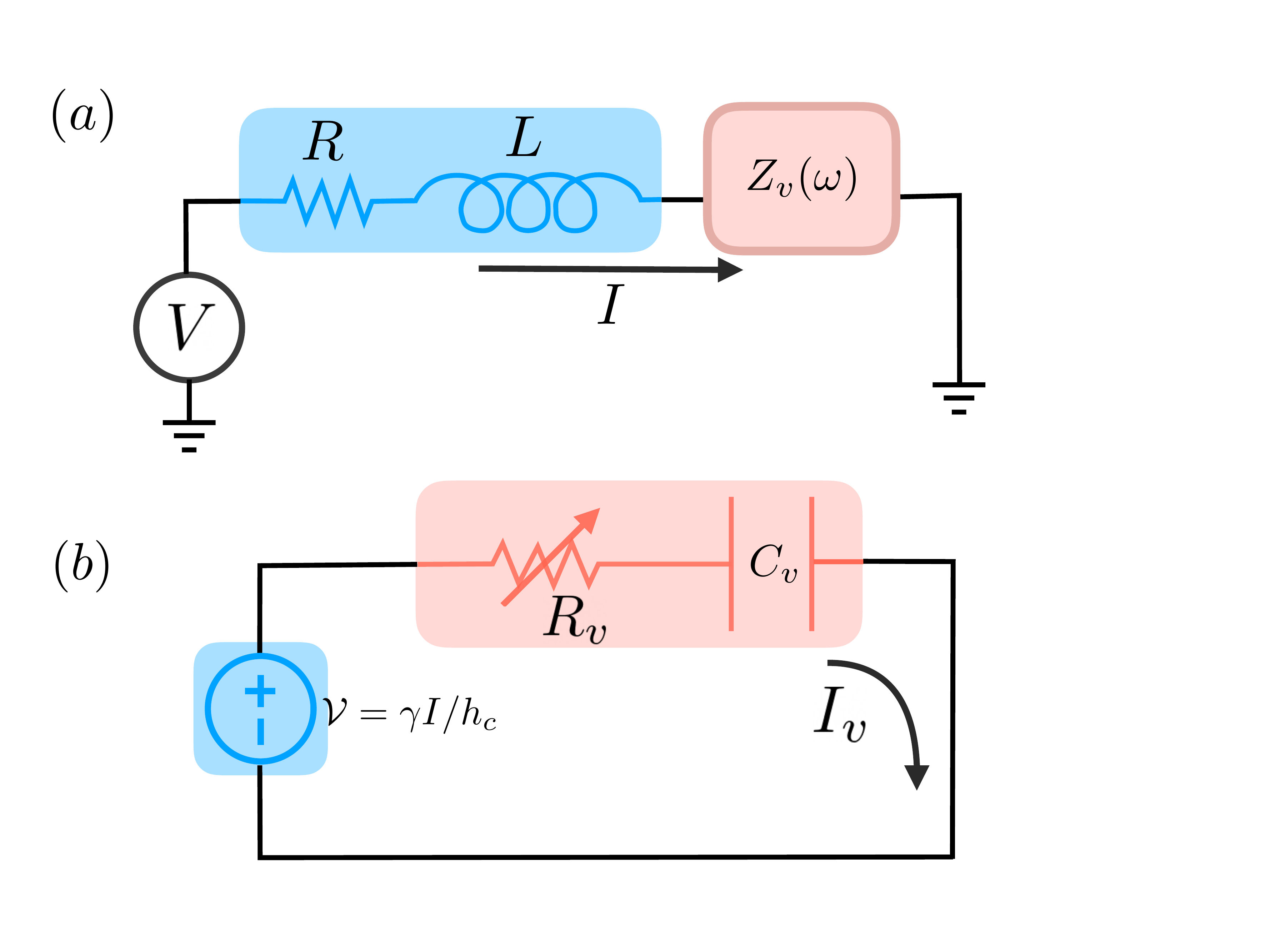}
\caption{The dynamics of the system in Fig.~\ref{fig1} can be described by two coupled circuits. (a) The electrical circuit, with a current $I$, resistance $R$, self-inductance $L$ (due to geometry), and effective impedance $Z_v(\omega)$ arising from the vortex-flow backaction on the electric circuit. Within the vortex circuit (b), the electric current $I$ acts as a bias $\mathcal{V}=\gamma I/h_c$ for the vortex flow, where $\gamma/h_c$ parametrizes the Magnus force between the electron and vortex degrees of freedom. Vortex flow through the magnetic bulk experiences resistance $R_v$, which is tunable by varying the temperature, whereas the accumulated winding texture builds up energy according to the capacitance $C_v$.}
\label{fig2}
\end{figure}

%%%%%%%%%%%%%%%%%%%%%%%%%%%%%%%%%%%%%%%%%%%%%%%%%%%%%%%%%%%%%%%%%%%%%%%%%
\sectionn{Main results.}|As we explain below, the dynamics of the system in Figs.~\ref{fig1}, \ref{fig3} can be understood by mapping to two coupled circuits, one for the electron flow and the other for the topological charge (vortex) flow. For the topological charge circuit [see Fig.~\ref{fig2}(b)], the electric current $I$ in the metal contact plays the role of a bias, which applies the vortex-motive force $\gamma I/h_c$, triggering a vortex current $I_v$. Here $\gamma/h_c$ is an interfacial spin-transfer torque parameter to be defined below.   The magnetic insulator itself behaves like a vortex capacitor ($C_v$) and resistor ($R_v$) in series.

For the electric circuit, a reciprocal electromotive force $\mathcal{E}_\text{EMF}=\gamma I_v/h_c$ arises from the coupling between electron and vortex dynamics~\cite{linearmomentum, PhysRevB.80.184411}, in series with the resistance $R$ and the geometric inductance $L$ of the metal contact. The Onsager reciprocity~\cite{PhysRev.37.405} between the two circuits can be expressed in the compact form $\vb{V}= \hat{\mathcal{R}}\vb{I}$,
\( \mqty(V\\ V_v)=\mqty(R+L\dv{}{t}  &  \gamma/h_c   \\ -\gamma/h_c & R_v)\mqty(I \\ I_v), \label{1} \)
where $V$ is the electric voltage drop across the metallic contact and $V_v=-\mathcal{Q}/C_v$ is the effective chemical potential associated with the accumulated topological charge $\mathcal{Q}\equiv\mathcal{N}$~\footnote{Note that the resistance matrix $\hat{\mathcal{R}}$ is antisymmetric since the metallic magnetization flips sign under time reversal (when one invokes Onsager reciprocity). One can also easily check the positive-definiteness of the dissipation $\mathbf{I}^\intercal \cdot \mathbf{V} \geq 0 $.}.
We will see that the electromotive force results in an impedance in the electric circuit, interpolating between a resistance (in the high-frequency response, compared with the characteristic $R_vC_v$ time) and an inductance (at low frequencies). As we discuss below, this inductance can be comparable with the geometrical inductance within the electric circuit, and, therefore, we propose to characterize the vorticity-current interaction through its contribution to the electrical rf inductance.
%This mapping between magnetic dynamics and a charging capacitor allows us to work within the electronic notation and calculate the stored energy using the standard notation for a capacitor. The energy of a capacitor is stored in the electric field or, in our case, the order parameter winding inside the magnet.

Lastly, by neglecting the geometric inductance $L$, allowing the device to charge for a time $\tau=R_vC_v$, and ignoring a small numerical term in the denominator, we  show that the charging efficiency, defined to be the ratio of the total energy stored to the total energy input, is 
\( \eta = \frac{1/2}{RR_vh_c^2/\gamma^2+1}, \label{2}  \)
from which we see explicitly that the efficiency benefits from a thinner metal contact $h_c$. The three parameters ($R,R_v$, and $\gamma/h_c$) correspond to three dissipation channels: electrical resistance of the metal, resistance of the vortex current in the magnetic insulator, and their mutual resistance, respectively.

%%%%%%%%%%%%%%%%%%%%%%%%%%%%%%%%%%%%%%%%%%%%%%%%%%%%%%%%%%%%%%%%%%%%%%%%%%%%%%%%%%%%%%%%%%%%%%%%%%%%%%%%%%%%%%%%%%%%%%%%%%%%%%%%%%%%%%%%%%%%%%%%%%%%%%%%%%%%%%%%%%%%%%%%%%%%%%%%%%%%%%%%%%%%%%%%%%%%%%%%%%%%%%%%%%%%%%%%%%%%%%%%%%%%%%%%%%%%%%%%%%%%%%%%%%%%%%%%%%%%%%%%%%%%%%
\sectionn{Biased vortex flow.}|A motive bias for vortex flow is established by a circulating electric current $I$ (see Fig.~\ref{fig1}) in a magnetically polarized metal contact ($\vb{M}= M\hat{\mathbf{z}}$). This current exerts a long-wavelength torque (per unit area) on the magnetic texture~\cite{jivortex},
\( \vb*{\tau}=\gamma\, n_z (\vec{j}\cdot \vec{\nabla})\vb{n},  \)
where $j=I/h_c\delta r$ is the electric current density and $n_z$ is the $z$ component of the order parameter. $\gamma\equiv \text{sign}(M)\, \pi \hbar h_{\text{eff}}/e$, where the lengthscale $h_{\text{eff}}$ can be loosely interpreted as the spatial extent of the torque, as a proximity effect at the interface, within the insulator. The torque does work upon magnetic dynamics at the rate
\( \dot{W}=\int dxdy \,\, \vb*{\tau}\cdot( \vb{n}\times \dot{\vb{n}})     = \gamma  \int dxdy \,\,  ( \vec{\mathcal{J}}\times \vec{j}) \cdot \hat{z}, \label{6}  \)
where the integration is performed over the interface. Here, we have taken, for simplicity, the magnitude of the order parameter to be fixed, $\vb{n}=1$. In the high-temperature regime, where $\vb{n}$ is fluctuating strongly, a similar result is expected, albeit with a renormalized prefactor. Eq.~(\ref{6}) indicates that the coupling between electron and vortex dynamics is Magnus cross-like (see Fig.~\ref{fig3}). In other words,  the electric current tangential to a magnetic interface produces a Magnus force on vortices, resulting in a transverse vortex flow. In reverse, by Onsager reciprocity, the vortex flow exerts an electromotive force on electrons within the metal leading to  a transverse electric current.  This underlies the mechanism for building up and  relaxing the winding texture in the magnetic insulator.

%%%%%%%%%%%%%%%%%%%%%%%%%%%%%%%%%%%%%%%%%%%%%%%%%%%%%%%%%%%%%%%%%%%%%%%%%%%%%%%%%%%%%%%%%%%%%%%%%%%%%%%%%%%%%%%%%%%%%%%%%%%%%%%%%%%%%%%%%%%%%%%%%%%%%%%%%%%%%%%%%%%%%%%%%%%%%%%%%%%%%%%%%%%%%%%%%%%%%%%%%%%%%%%%%%%%%%%%%%%%%%%%%%%%%%%%%%%%%%%%%%%%%%%%%%%%%%%%%%%%%%%%%%%%%%
\sectionn{Mapping onto two coupled circuits.}|We first consider the vortex dynamics in the magnetic insulator, by exploiting the duality between the $XY$ magnet and electrostatics in two dimensions~\cite{Kosterlitz_1974, shuduality}. In the low-temperature regime (for simplicity) $\vb{n}$ is in plane and has a fixed magnitude (though, the results we obtain also generalize to the high-temperature regime where the magnitude of $\vb{n}$ is allowed to fluctuate).   The duality is accomplished by identifying the total winding of the magnetic order parameter with the electric charge $\mathcal{Q}=\mathcal{N}$, and the spatial gradients of the order-parameter angle with the electric field $\vec{E}=\mathcal{A}\vec{\nabla}\varphi\times \hat{z}$, where $\mathcal{A}$ is the order-parameter stiffness whose magnitude is on the order of $J/a$ ($J$ is the exchange energy and $a$ is the lattice spacing). 
We can now recast the definition of the winding number (\ref{4}) as  Gauss's law for the electric charge $\int d\vec{s}\cdot \vec{E}=\mathcal{Q}/\epsilon$, where $d\vec{s}=d\vec{l}\times \hat{z}$ is the line element in the azimuthal direction and $\epsilon=1/\mathcal{A}$ is the permittivity. Note that making this identification requires a rescaling of the position variables with the mapping $\vb{r} \rightarrow \vb{r}/2\pi$ such that both sides of the equation for Gauss's law have equal units. Mapping the energy expression for the insulating magnet to electrostatic notation, gives                    
\( \mathcal{E} =h_m \mathcal{A}  \int dxdy\,\, (\nabla \varphi)^2= h_m  \int dxdy\,\, \frac{\epsilon\, \vec{E}^2 }{2} ,  \label{7}\)
where $h_m$ is the height of the magnetic insulator.

Therefore, driving  topological charges (vortices) from the inner edge to the outer edge can be interpreted as a charging-capacitor process, which is triggered by a charge transfer (that is linked to the winding number) across the annulus. Noting that the power (\ref{6}) can be rewritten as $\dot{W}=\gamma I I_v/h_c$, we can view the metallic contact as a battery with voltage $\mathcal{V}= \gamma I/h_c$ acting on a vortex $R_vC_v$ circuit, as illustrated in Fig.~\ref{fig2}(a). 
% Therefore, all discussions above conspire to yield the capacitor circuit shown in Fig. \ref{fig2}(a) and written in Eq. (\ref{1}) for vortex dynamics. 
The effective capacitance can be extracted by simply  equating the energy (\ref{7}) with $\mathcal{E}=\mathcal{Q}^2/2C_v$, whereas Fick's law~\cite{PhysicalKinetics} 
$\vec{\mathcal{J}}=-\sigma_v \vec{\nabla}\mu$
gives the resistance $R_v = \Delta \mu /I_v$,
where $I_v=2\pi r \mathcal{J}$ is the vortex current and $\Delta \mu$ is the motive force on the vortex flow. Thus, we arrive at the effective vortex capacitance  and  resistance
\( C_v=\frac{1}{\mathcal{A}} \frac{2\pi r}{ h_m \delta r},\,\,\,\,\, R_v= \frac{1}{\sigma_v} \frac{\delta r}{2\pi r}. \label{9}\)
Here $\sigma_v^{-1}$ is the vortex resistivity whose main contributions arise from vortex collisions (such as umklapp scattering, disorder, etc) and Gilbert damping.

The vortex current acts reciprocally on the electric circuit, as summarized in Eq.~(\ref{1}), from which we wish to determine the total impedance acting on the electric current. After Fourier transforming and solving for the electric response by eliminating the vortex current, we arrive at the total impedance:
\(Z(\omega)\equiv\frac{V(\omega)}{I(\omega)} = R+i\omega L+ \frac{i\omega C_v \gamma^2 /h_c^2}{1+i\omega R_vC_v},  \)
where the last term [henceforth denoted $Z_v(\omega)$] is the vorticity impedance, arising from the coupling between electron and vortex dynamics. In the high frequency regime ($\omega\gg 1/\tau$) where $\tau=R_vC_v$ is the time scale of the vortex charging (or discharging) process, one obtains $Z_v(\omega)=\gamma^2/h_c^2R_v$, indicating that the magnetic insulator, generating an electromotive force against the input electric current, behaves like a resistor in the electric circuit. In the opposite regime where $\omega\ll 1/\tau$, we have $Z_v(\omega)=i\omega C_v\gamma^2/h_c^2$, suggesting that the magnetic insulator plays the role of an inductor with $L_v=C_v\gamma^2/h_c^2$.

%The effect of geometrical inductance $L$ can be neglected because it is orders of magnitude smaller than $L_v$~\footnote{The ratio of these inductances is $\frac{L_v}{L}\sim \frac{1}{\alpha^2} \frac{e^2}{aJ} \frac{a^2}{S_m},$ where $\alpha\equiv e^2/\hbar c$ is the fine structure constant revealing that $L$ is a relativistic effect.}.

%%%%%%%%%%%%%%%%%%%%%%%%%%%%%%%%%%%%%%%%%%%%%%%%%%%%%%%%%%%%%%%%%%%%%%%%%%%%%%%%%%%%%%%%%%%%%%%%%%%%%%%%%%%%%%%%%%%%%%%%%%%%%%%%%%%%%%%%%%%%%%%%%%%%%%%%%%%%%%%%%%%%%%%%%%%%%%%%%%%%%%%%%%%%%%%%%%%%%%%%%%%%%%%%%%%%%%%%%%%%%%%%%%%%%%%%%%%%%%%%%%%%%%%%%%%%%%%%%%%%%%%%%%%%%%
\sectionn{Battery efficiency and quantitative estimates.}|  The dc electric current $I$ flowing in the metal contact [Fig.~\ref{fig1}] eventually results in a steady-state magnetic texture with winding density $ \partial_x \varphi = \gamma I /\mathcal{A} h_ch_m\delta r$, and an associated free energy
\( \mathcal{E}=\frac{1}{2}C_v\mathcal{V}^2= \frac{1}{\mathcal{A}} \frac{\pi r}{h_m\delta r} \Big( \frac{\gamma I}{h_c} \Big)^2,   \)
at time $t\gg \tau=R_vC_v=1/\mathcal{A}\sigma_vh_m$. Here, the vortex conductivity $\sigma_v=\rho_v\mu =\rho_v D/k_BT$ depends on the temperature through the vortex density $\rho_v$ and vortex mobility $\mu$ (which is related to the diffusion constant $D$ by the Einstein relation). In the extreme limit $T\ll T_c$, where $\rho_v\sim 0$, we have zero vortex conductivity leading to $\tau \rightarrow \infty$.   In the opposite regime (near the Curie temperature $T_c\sim J/k_B$), the order parameter varies on the atomic scale, $\rho_v\sim 1/a^2$ and $D\sim Ja^2/\hbar$, giving the lower bound of the charging time $\tau \sim \hbar/J$. Thus, the vortex conductivity $\sigma_v$ and $\tau$ are highly tunable by temperature.

To obtain the efficiency $\eta$ of the charging process, we neglect the geometrical inductance of the metal contact and allow the device to charge for a time $\tau$. The charging will be accomplished by using a single square wave pulse of current $I$. The total external energy input during the charging process is 
\( \mathcal{W}=\int_0^\tau dt \,\, I \,V(t)=I^2R\tau +\tau\frac{\mathcal{V}^2}{R_v}(1-e^{-1}),   \)
where $V(t)$ is the electric voltage drop across the metal contact that can be obtained by solving Eq.~(\ref{1}). These terms take into account the energy loss due to Joule heating and vortex motion as well as the stored energy within the magnetic texture. By dropping the numerical factor in the second term which is of order unity and depends on the details of the charging process, the efficiency of the charging-process becomes 
\( \eta= \frac{\mathcal{E}}{\mathcal{W}}= \frac{1/2}{RR_v h_c^2 /\gamma^2 +1}.  \)
Considering the regime where $\tau\sim \hbar/J$, we have
\( \frac{RR_vh_c^2}{\gamma^2}\sim \frac{h_mh_c}{h_{\text{eff}}^2} \frac{1/k_F^2}{a\,l },  \)
where $l$ and $k_F$ are the mean free path and Fermi wavelength of electrons within the metal, respectively. It is clear that the efficiency benefits from improving the conducting quality of the metal and  decreasing  thicknesses of both insulating magnet and   metallic contact, which makes sense intuitively. Taking the geometrical inductance $L$ into account, we should also obtain a better efficiency. In the limiting case of $L\rightarrow \infty$, where the charging process is adiabatic, the efficiency can, in principle, approach $1$.

The maximal energy-storage capacity is another quantity of interest. This is dictated by the Landau criterion for  energetic stability~\cite{Sonin:2010aa}, where the magnetic texture is maximally wound. It is achieved when the winding texture energy [$\propto \mathcal{A}(\partial_x\varphi)^2$] is comparable to the easy-plane anisotropy energy ($\propto \mathcal{K}$) that fixes the winding within the easy-plane. Let us take the bulk stiffness to be $\mathcal{A}= 5\times 10^{-12}~\text{J/m}$, an easy-plane anisotropy strength of $\mathcal{K}=5\times10^{5}~\text{J}/\text{m}^3$,  and mass density $5.11~\text{g/cm}^3$ (yttrium iron garnet), which yields for the winding density $1/\partial_x\varphi = \sqrt{\mathcal{A}/\mathcal{K}}\sim 3\, \text{nm}$ and a specific energy density of $100~\text{J/kg}$.   Such an energy can be loaded by applying a electric current density of $ 10^{12} \text{A/m}^2$ within a thin metal contact, which is  feasible experimentally \cite{Parkin190}. We can further increase the specific energy density by enhancing the easy-plane anisotropy. For example, in the extreme limit where the order parameter can vary on the atomic scale, $1/\partial_x\varphi\sim a$, we have the  specific energy density $10^4~\text{J/kg}$, which is about an order of magnitude below the capacity of lithium-ion batteries.

To characterize the vorticity-current interaction, which underlies the mechanism of our proposal, we make the suggestion to measure its contribution to the electric inductance in the rf response. To this end, we note that $L_v$ can be manufactured to be comparable with the geometrical inductance $L$:
\( \frac{L_v}{L}\sim \frac{1}{\alpha^2} \frac{e^2/a}{J}\frac{a^2}{h_m \delta r}\sim 1,  \)
 where $\alpha$ is the fine structure constant and we have used $\delta r h_m\sim 100\, \text{nm}^2$. Alternatively, one can measure the (transient) vortex discharging process, where the electric voltage of the metal is $V(t)=V_{\text{max}}e^{-t/\tau}$, by solving Eq.~(\ref{1}) with an open electric circuit. For a thin contact ($h_c\sim h_{\text{eff}}$), one obtains that 
  \(  V_{\text{max}}\tau\sim \frac{\hbar}{e} \frac{r}{\sqrt{\mathcal{A}/\mathcal{K}}}.  \)
 Assuming $r\sim 1~\mu$m and $\tau\sim~10\, \text{ns}$ which should be easily accessed experimentally, we get a measurable voltage drop of $V_{\text{max}}\sim 10^{-4}$~V.

%%%%%%%%%%%%%%%%%%%%%%%%%%%%%%%%%%%%%%%%%%%%%%%%%%%%%%%%%%%%%%%%%%%%%%%%%%%%%%%%%%%%%%%%%%%%%%%%%%%%%%%%%%%%%%%%%%%%%%%%%%%%%%%%%%%%%%%%%%%%%%%%%%%%%%%%%%%%%%%%%%%%%%%%%%%%%%%%%%%%%%%%%%%%%%%%%%%%%%%%%%%%%%%%%%%%%%%%%%%%%%%%%%%%%%%%%%%%%%%%%%%%%%%%%%%%%%%%%%%%%%%%%%%%%%
\sectionn{Summary and outlook.}| We have proposed an experimentally feasible energy storage concept in insulating magnets based on the collective transport of vortices, emerging out of the topologically nontrivial real-space order-parameter textures. This allows to utilize the current-magnet interaction with a focus on the dynamics of topological textures rather than the conventional spin currents. The energy associated with the winding texture can be loaded by electric means  which biases a vortex flow within the magnet~\cite{jivortex, quantumvortex}. The system is mapped  onto two coupled circuits, where we interpret the energy-loading process as a capacitor-charging action. This energy storage is attractive due its potential longevity~\cite{Smith_2010, 7555318}, endowed by the topological nature of the vorticity, and its compatibility with integrated spintronic circuits~\cite{Allwood1688, Chumak:2015aa, Khitun_2010, PhysRevX.5.041049} and quantum-information processes based on insulating magnets~\cite{PhysRevB.95.144402,PhysRevB.101.014416,PhysRevB.100.174407}.

One could envision a variety of  generalizations of our proposal by exploiting different topological hydrodynamics. An immediate example is the magnetic hedgehog in three dimensions. When a hedgehog passes through a chiral magnet~\cite{Milde1076}, a finite skyrmion density is built up which is associated with finite energy and can be devised to store energy. The resultant skyrmion density is protected by Dzyaloshinskii-Moriya interaction which plays a role of easy-plane anisotropy for winding texture. We remark that this is the generic property of $n$-dimensional nonlocal topological defects, which would establish $(n-1)$-dimensional nonlinear textures when  they flow through a medium, dictated by the generalized Stokes' theorem.  Other types of topologically conserved local defects, such as skyrmions in two-dimensional magnetic films~\cite{Hectorskyrmion} and three-dimensional skyrmionic textures in frustrated magnets~\cite{PhysRevB.100.054426}, can also be quite valuable  potentially for energy-storage purpose. All these possibilities provide opportunities to explore energy storage concepts based on spin degrees of freedom and deserve further investigation.

We are grateful to Mostafa Ahari, Jiang Xiao, and Wei Han for insightful discussions. The work was supported by the U.S. Department of Energy, Office of Basic Energy Sciences under Award No. DE-SC0012190.

\end{document}